\documentclass[aps,twocolumn,floatfix,footinbib]{revtex4-1}
\usepackage{amsmath}
\usepackage{amsfonts}
\usepackage{graphicx}
\usepackage{color}
\usepackage{bm}
\usepackage{hyperref}

\newcommand{\TTT}{\mathcal{T}}

\newcommand{\bbone}{{\rm 1\hspace*{-0.4ex}\rule{0.1ex}{1.52ex}\hspace*{0.2ex}}}

\begin{document}
\title{Braiding properties of Majorana Kramers Pairs}
\author{Konrad W\"olms,$^1$ Ady Stern,$^2$ and Karsten Flensberg$^1$}
\affiliation{$^1$Center for Quantum Devices, Niels Bohr Institute,
University of Copenhagen, Universitetsparken 5, 2100 Copenhagen, Denmark\\
$^2$Department of Condensed Matter Physics, Weizmann Institute of Science, Rehovot 76100, Israel}

\date{\today}

\begin{abstract}
We consider the braiding of Kramers pairs of Majorana bound states. We derive
the most general transformation on the many-body ground state that is applied as
the result of such a braiding process. The result is derived in the context of a
simple toy model, but we will show that it has the most general form that is
compatible with local and global conservation of electron parity. In accordance
with earlier work the resulting transformation turns out to be path dependent,
which shows that Kramers pairs of Majorana bound states cannot be used for
topological quantum computation. We also discuss under which conditions the
result is path independent and corresponds to two independent exchanges of pairs
of Majorana bound states.
\end{abstract}

\maketitle

For certain classes of topological superconductors (D, BDI), Majorana bound
states (MBS) appear at topological phase boundaries or in vortices. Those MBS
are a signature of the non-trivial topological phases and they can be
used for topological quantum computation \cite{nayak_non-abelian_2008}. As such
MBS have received a lot of attention, both from the theoretical and the
experimental side
\cite{beenakker_search_2013,leijnse_introduction_2012,alicea_majorana_2010}.

Another closely related topological class is DIII. This class has time reversal
symmetry that squares to minus one and exhibits Kramers pairs of MBS  at
topological phase boundaries or in vortices. Even though there have not been
experimental attempts yet to realize this topological phase, there have been
many theoretical proposals on how to obtain this
phase\cite{wong_majorana_2012,zhang_time-reversal-invariant_2013,nakosai_majorana_2013,nakosai_topological_2012,deng_majorana_2012,keselman_inducing_2013,gaidamauskas_majorana_2014,haim_time-reversal-invariant_2014,danon_interaction_2015}.
A natural question to ask is whether those Kramers pairs of MBS can be used for
topological quantum computation similar to MBS appearing in classes D and BDI.
This question is equivalent to the questions whether Kramers pairs of MBS are
non-Abelian anyons and whether their adiabatic exchange generates a
transformation that is independent of the details of the exchange path.

In earlier work we argued that this cannot be the case because Kramers pairs of
MBS have a local degree of freedom that can be manipulated
adiabatically\cite{wolms_local_2014}. This makes any exchange of two such
Kramers pairs path dependent, which means that they are not anyons or
equivalently that they cannot be used for topological quantum computation.

In this article, we study the general transformation that occurs when braiding
two Kramers pairs of MBS. We study it in the context of a simple toy model, but
this will not limit the generality of the results because even in this simple
model we find the most general transformation that is allowed by local parity
conservation.

\begin{figure}
    \centering
    \includegraphics[]{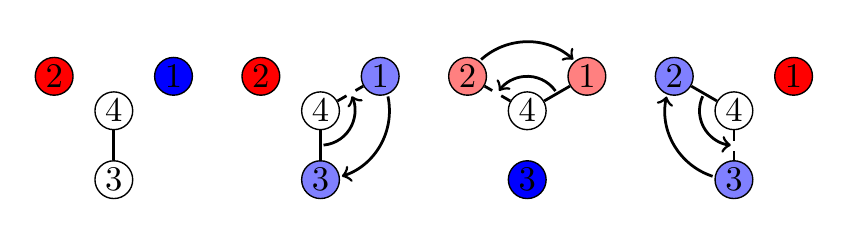}
    \caption{Each circle denotes a site with a Kramers pair of Majorana fermions. Lines between the
    sites denote couplings. As a single coupling is switched from position to the next, the
    corresponding zero energy Kramers pair (color)  moves the opposite direction. Note that the sites label
    the static basis and therefore the numbers don't change.}
    \label{fig:toymodel}
\end{figure}

The toy model we study is the time-reversal-symmetric analog of braiding by
tuning couplings \cite{sau_controlling_2011}. It consists out of four Kramers
pairs of Majorana fermions $\chi_i$, $\tilde\chi_i=\TTT\chi_i\TTT^{-1}$, which we
write as a vector $\bm{\chi}_i=(\chi_i,\tilde \chi_i)^T$. The most general
time-reversal-symmetric coupling between two Majorana Kramers pairs is
\begin{equation}
    H_{jk}=it\bm{\chi}_j^T\sigma_ze^{i\beta_{jk}\sigma_y}\bm{\chi}_k.
    \label{eqn:generalKramersPairCoupling}
\end{equation}
In the toy model there are only three coupling terms, $H_{41}$, $H_{42}$ and
$H_{43}$. In order to have the appropriate ground state degeneracy at all times,
there is at least one non-zero coupling and at most two at each point in
parameter space. Furthermore, we take the Hamiltonian with two couplings to be
of the form
\begin{equation}
H=\cos\theta H_{ij} +\sin\theta H_{ik},
    \label{eqn:switchingHamiltonian}
\end{equation}
such that $\theta$ controls the relative strength of the couplings.  The effect
of switching from one coupling to another, by means of tuning $\theta$ from 0 to
$\pi/2$, is to move a zero energy Kramers pair from one site to another. The
situation is depicted in Fig. \ref{fig:toymodel}.

In order to study the adiabatic braiding process we have to specify a basis at
each point in parameter space. At each such  point we have two zero energy
Kramers pairs of Majorana bound states, which we denote with $\bm{{X}}_1$ and
$\bm{{X}}_2$. We assume that those are local and initially start out as
$\bm{{X}}_1 \sim \bm{\chi}_1$, $\bm{ {X}}_2 \sim\bm{\chi}_2$, where $\sim$ means
up to (generally different) local rotations $e^{i\alpha\sigma_y}$. We construct
a local basis by forming local fermions $D_{\eta} = \tfrac 1 2 ( {X}_\eta +
i\tilde{ {X}}_\eta)$, where we used a Greek index which is either 1 or 2 to
distinguish it from the Latin indices which run from 1 to 4. The basis we use at
each point in parameter space is now given as
\begin{equation}
    \begin{split}
        |00\rangle &
        \\
        |01\rangle&=D_2^\dagger |00\rangle,
        \\
        |10\rangle&=D_1^\dagger |00\rangle,
        \\
        |11\rangle&=D_1^\dagger D_2^\dagger|00\rangle.
        \end{split}
    \label{eqn:instantaneousBasis}
\end{equation}

Note that there is a problem with our local basis choice
\eqref{eqn:instantaneousBasis} when completing one braid. By definition our
$\bm{ {X}}_\eta$ are local and after the braiding is done they'll end up as $\bm
{X}_1\sim\bm\chi_2$ and $\bm {X}_2 \sim \bm\chi_1$. This means that our local
basis \eqref{eqn:instantaneousBasis} does not go back to itself after completing
a loop in parameter space. We can correct for that by an explicit basis
transformation at the end. This way the unitary transformation, $U$, due to
braiding, factors into local Berry phases, $U_{\text{local}}$, followed by a
global basis transformation, $B$. We have
\begin{equation}
    U = B U_{\text{local}}.
    \label{eqn:braidingTransformationDecomposition}
\end{equation}
In principle one could perform a basis transformation that exchanges $\bm {X}_1$
and $\bm {X}_2$ at an earlier point during the exchange. In that case one still
has to make sure that the phases of the states at the end match up exactly with
the ones in the beginning, otherwise one needs to do another basis
transformation. Therefore it is most convenient to perform everything as a
single bases transformation, which includes potential phases, at the end of the
process.

In our earlier work \cite{wolms_local_2014} we studied $U_{\text{local}}$. It
generally takes the form
\begin{equation}
    U_{\text{local}}=e^{\frac {\varphi_1} 2 {X}_1\tilde {X}_1}e^{\frac {\varphi_2} 2
 {X}_2\tilde {X}_2},
    \label{eqn:generalLocalTransformation}
\end{equation}
where the $\varphi_i$s are calculated as
\begin{equation}
    \varphi_\eta = \frac 1 2 \int_{\mathcal{C}}
    \{ {X}_\eta,\nabla_{\bm\lambda}\tilde {X}_\eta\}\cdot\mathrm{d}\bm\lambda,
    \label{eqn:mixingPhase}
\end{equation}
with $\mathcal{C}$ being the path in parameter space during the exchange.  We
refer to this as mixing by an angle $\varphi_1$ or $\varphi_2$ of the Kramers
pair 1 and 2 respectively. Note that the integration path is not closed.
Therefore a gauge transformation will change \eqref{eqn:mixingPhase} according
to the initial and final gauge choice. We will discuss this issue in detail
later.  In this work, we will focus on finding the basis transformation $B$ in
order to calculate \eqref{eqn:braidingTransformationDecomposition}.

To find the basis transformation we have to determine the final form of $\bm
{X}_\eta$. The transformation $B$ is then determined as the transformation that fulfills
\begin{equation}
    B \bm {X}_\eta^{\text{final}}B^\dagger=\bm {X}_\eta^{\text{initial}}.
    \label{eqn:BasisChangeDefinition}
\end{equation}
This way $B$ is, of course, only defined up to an abelian phase, but this phase
will be strongly system dependent in any case and we do not consider it. From
the definition it is also obvious that $B$ depends on the gauge choice at the
initial and final position. We will discuss this dependence in detail below.

\begin{figure}
    \centering
    \includegraphics[]{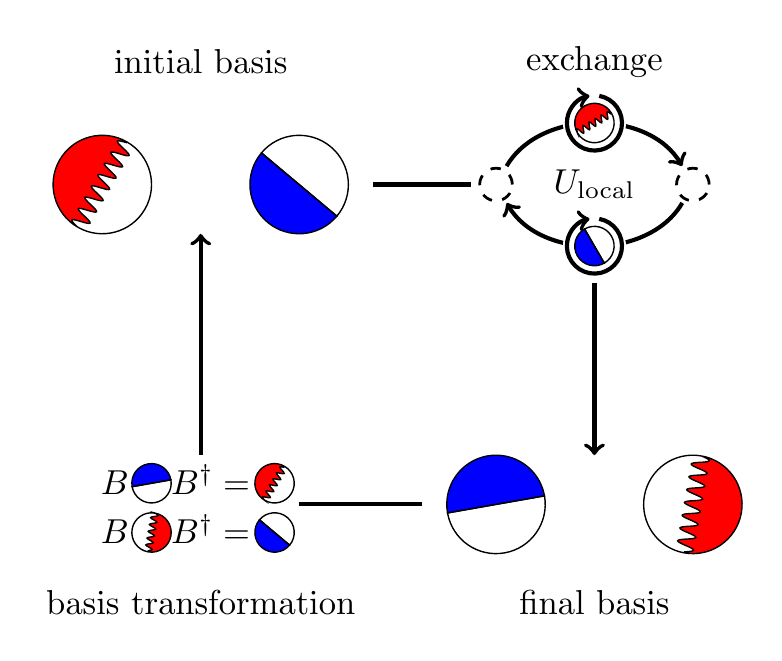}
    \caption{Graphical representation of \eqref{eqn:braidingTransformationDecomposition}. A loop in
    parameter space is achieved by a continuous exchange followed by a basis transformation. The
    circles represent Kramers pairs of Majorana fermions. The ambiguity of dividing a circle into two
    represents the gauge freedom in choosing Kramers partners. Two different lines and colors are used to
    divide the circles to distinguish the two Kramers pairs.}
    \label{fig:decomposition}
\end{figure}

A pictorial representation of the  decomposition
\eqref{eqn:braidingTransformationDecomposition} is given in figure
\ref{fig:decomposition}, where the circles represent Kramers pairs and the gauge
freedom in the choice of Kramers partners is represented by choice of angle at
which the circles are split into two halves.

One might be worried about the transformation properties of equation
\eqref{eqn:braidingTransformationDecomposition} under gauge transformations.
First of all it is clear that the decomposition can only hold as long as we
restrict ourselves to local gauge transformations. Under a gauge transformation
$W(\bm\lambda)$ the operator $U$ should transform as
\begin{equation}
    U\to W(\bm\lambda_{\text{initial}})UW^\dagger(\bm\lambda_{\text{initial}}),
    \label{eqn:generalGaugeTransformation}
\end{equation}
which means that the transformation $U$ is basis dependent as one would expect.
We explicitly show that this relation holds for our decomposition
\eqref{eqn:braidingTransformationDecomposition} under local gauge
transformations, which have the form
\begin{equation}
    W_{\text{local}} =
    e^{f_1(\bm\lambda) {X}_1\tilde {X}_1}e^{f_2(\bm\lambda) {X}_2\tilde {X}_2}.
    \label{eqn:localGaugeTransformation}
\end{equation}
This is important, because we will show below that $B$ and $U_{\text{local}}$ do
not satisfy the same transformation property individually.

We denote $W_{\text{local}}(\bm{\lambda}_{\text{initial/final}})$ by
$W_{\text{i/f}}$ respectively.  It follows directly from $B\bm
{X}_\eta^{\text{final}}B^\dagger =\bm {X}_\eta^{\text{initial}}$ that
$W^{\phantom{\dagger}}_{\text{i}}B_{\phantom{\text{i}}} W^\dagger_{\text{f}}
W^{\phantom{\dagger}}_{\text{f}}\bm
{X}_\eta^{\text{final}}W^\dagger_{\text{f}}W^{\phantom{\dagger}}_{\text{f}}
B^\dagger W^\dagger_{\text{i}}=W^{\phantom{\dagger}}_{\text{i}}\bm
{X}_\eta^{\text{initial}}W^\dagger_{\text{i}}$.  Hence $B$ transforms as $B\to
W_{\text{i}} B W_{\text{f}}^\dagger$. If one replaces $ {X}_\eta, \tilde
{X}_\eta$ by $W_{\text{local}} {X}_\eta W^\dagger_{\text{local}}, W^\dagger
_{\text{local}}\tilde  {X}_\eta W_{\text{local}}$ in \eqref{eqn:mixingPhase} one
easily finds that the local phases transform as $\varphi_\eta\to\varphi_\eta +
(2f_{\eta}(\bm\lambda_{\text{final}})- 2f_{\eta}(\bm\lambda_{\text{initial}})$),
from which it follows that $U_{\text{local}}$ transforms as $U_{\text{local}}\to
W^{\phantom{\dagger}}_{\text{f}}U_{\text{local}}W^\dagger_{\text{i}}$. Hence
neither $B$ nor $U_{\text{local}}$ transform according to
\eqref{eqn:generalGaugeTransformation}, in particular because they depend on the
gauge choice at $\bm\lambda_{\text{final}}$. This dependence however drops out
of their product, such that $U$ indeed fulfills
\eqref{eqn:generalGaugeTransformation}, proving  that our decomposition
\eqref{eqn:braidingTransformationDecomposition} is valid for any local gauge
choice.

Before we continue and calculate B let us comment on a particular interesting
local gauge choice.  This gauge choice is defined by
$U_{\text{local}}=\bbone$, which is always possible. This way one simply has
$U=B$, which is useful for numerical calculations, because in this way the
calculation of the transformation does not require taking derivatives. A
discretization of a path in parameter space only needs to be fine grained enough
in order to fix the gauge choice.

To calculate $B$ we have to find $\bm {X}_\eta^{\text{final}}$ and therefore we
need to understand how $\bm {X}_\eta$ changes during one switching process
$\eqref{eqn:switchingHamiltonian}$ when taking $\theta=0\to\tfrac \pi 2$.  We
require locality of the zero energy Kramers pairs. Therefore one of the $\bm
{X}_\eta$ will stay constant through the switching process (up to parameter
dependent local gauge choices). The other of the $\bm {X}_\eta$ will move
between sites and by solving $[H,\bm{ {X}}_{\eta}]=0$ we find
\begin{equation}
    \bm{{X}}_\eta = e^{i\alpha\sigma_y}\left(\cos \theta e^{i\beta_{ij}\sigma_y}\bm{\chi}_j - \sin \theta
    e^{i\beta_{ik}\sigma_y}\bm{\chi}_k\right),
    \label{switchingInstantaneousMajorana}
\end{equation}
where $\alpha$ is an arbitrary gauge choice that may depend on the $\beta$s and
$\theta$.

We want to patch three switching processes together such that $\bm {X}_\eta$ is
continuous. In order to do that conveniently we pick a particular parameter
dependence for $\alpha$, such that $\bm {X}_{\eta}(\theta=0)=\pm\bm\chi_{j}$ and
$\bm {X}_{\eta}(\theta=\tfrac \pi 2)=\mp \bm\chi_k$. The sign change is
motivated by analogy to switching processes of single Majorana fermions in class
D systems. A particular gauge  choice of $\alpha$ that gives the desired $\bm
{X}_\eta$ at $\theta=0,\tfrac \pi 2$ is
$\alpha_{\pm}=-\beta_{ij}\cos\theta-\beta_{ik}\sin\theta + (1\mp1)\tfrac \pi 2$.

Table \ref{tab:braidingTable} shows $\bm {X}_{\eta}$ initially and after each
switching process.  It has exactly the same structure as for single, class D,
MBS. As the transformation from initial to final Kramers pairs of MBS we get
\begin{equation}
    \begin{split}
        \bm {X}_1^{\text{initial}}=\bm{\chi}_1 & \to
        \bm {X}_1^{\text{final}}=\bm{\chi}_2, \\
        \bm {X}_2^{\text{initial}}=\bm{\chi}_2 & \to
        \bm {X}_2^{\text{final}}=-\bm{\chi}_1.
    \end{split}
    \label{eqn:majoranaFinalInitial}
\end{equation}
Therefore the basis transformation that generates this transformation is simply
\begin{equation}
    B = e^{\frac \pi 4(\chi_1\chi_2 + \tilde{\chi}_1\tilde{\chi}_2)},
    \label{eqn:generalAfterBraidingBasisTransformation}
\end{equation}
which is structurally the same as two independent braiding transformations of
pairs of MBS . According to \eqref{eqn:braidingTransformationDecomposition} the
total braiding transformation is
\begin{equation}
    U = e^{\frac \pi 4(\chi_1\chi_2 + \tilde{\chi}_1\tilde{\chi}_2)}
    e^{\frac {\varphi_1} 2 \chi_1 \tilde{\chi}_1+ \frac {\varphi_2} 2 \chi_2\tilde{\chi}_2}.
    \label{eqn:generalU}
\end{equation}
\begin{table}
    \begin{center}
        \begin{tabular}{lcccc}
            \\
            \hline
            & $H_{43}$ & $H_{41}$ & $H_{42}$ & $H_{43}$  \\
            \hline
            $\bm{ {X}}_1$  & $\bm{\chi}_1$ &
            $-\bm{\chi}_3$& $-\bm{\chi}_3$
            &  $\bm{\chi}_2$ \\
            $\bm{ {X}}_2$  & $\bm{\chi}_2$ &
            $\bm{\chi}_2$& $-\bm{\chi}_1$
            &  $-\bm{\chi}_1$\\
            \hline
        \end{tabular}
    \end{center}
    \caption{Instantaneous $\bm  {X}$ for a braiding process requiring them to be continuous. The
        process is of the form that is depicted in figure \ref{fig:toymodel}.}
    \label{tab:braidingTable}
\end{table}

The question arises how general this transformation is, because we obtained it
in the context of our simple toy model. From the construction it is clear that
we can always choose a parameter dependent basis such that Eqs.
\eqref{eqn:majoranaFinalInitial} and
\eqref{eqn:generalAfterBraidingBasisTransformation} are true. Therefore the
question reduces to whether $U_\text{local}$ can be more complicated than
\eqref{eqn:generalLocalTransformation}. Since we assume that both Kramers pairs
of Majorana fermions are always decoupled, the local parity operators $i
{X}_\eta\tilde {X}_\eta$ are conserved quantities. The only local
transformations which we can construct out of those are given by
$U_\text{local}$. Hence our toy model already describes the most general
possible braiding transformation for Kramers pairs of MBS.

Because the phases in Eq. \eqref{eqn:generalLocalTransformation} are path
dependent, the exchange of Kramers pairs of MBS cannot be used for topological
quantum computation. The natural question arises whether there are additional
conditions under which the phases become path independent. We will show that
this is the case in the absence of local mixing.  By local mixing we mean the
transformations that can arise when changing the parameters along any closed
trajectory in parameter space that does not braid $\bm {X}_1,\bm {X}_2$.
The transformation describing this process will simply be $U=U_{\text{local}}$,
with the difference that the integral in Eq. \eqref{eqn:mixingPhase} is over a
closed trajectory in this case. The absence of local mixing means that
$\varphi_\eta$ is zero for any such closed loop. This result is useful because
the presence or absence of local mixing is a property of individual Kramers
pairs of MBS, but we can use it to make a statement about braiding of several
Kramers pairs.

Before we discuss the implications of the presence or absence of local mixing,
note that if the phases in Eq. \eqref{eqn:generalLocalTransformation} are path
independent, then there cannot be local mixing, because otherwise we could
always add a closed path with non-vanishing mixing phase to the open path in Eq.
\eqref{eqn:mixingPhase}, violating the path independence of the phase.  The
absence of local mixing is therefore necessary in order to have constant phases
in Eq.  \eqref{eqn:generalLocalTransformation}. We now proceed to show that it
is also sufficient and moreover that in the absence of local mixing the braiding
transformation can be reduced to two independent braidings of $\chi_1,\chi_2$
and $\tilde\chi_1,\tilde\chi_2$ respectively.

\begin{figure}
    \centering
    \includegraphics{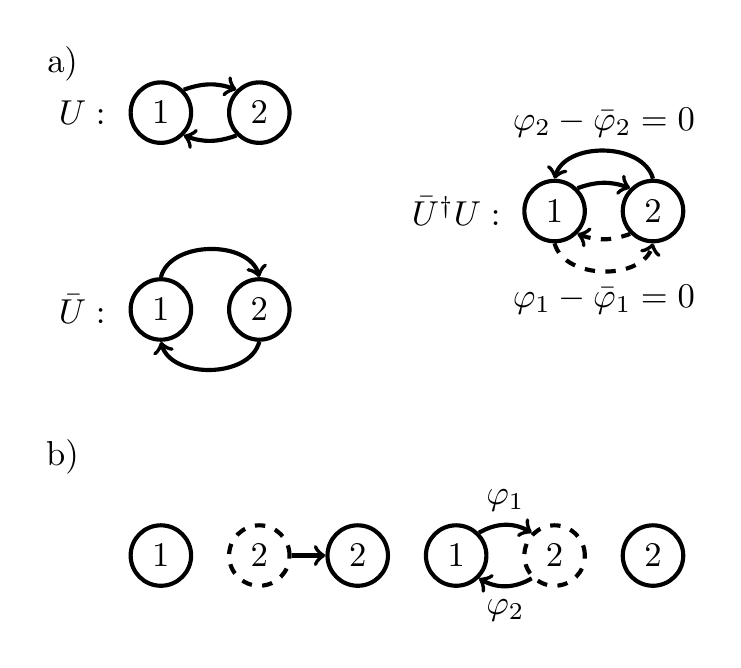}
    \caption{(a) Two exchanges along different paths can be combined into two
        local operation (solid and dashed line). Consequently the mixing angles
        are path independent if local mixing angles are zero. (b) One Kramers
        pair gets moved out of the way for the other to make a loop.  During the
        first half pair 1 rotates by $\varphi_1$ and during the second half it
        rotates by $\varphi_2$.  If there is no local mixing this implies
        $\varphi_1+\varphi_2=0$.}
    \label{fig:signargument}
\end{figure}

We first show that $\varphi_\eta$ are path independent. Assume we have two
braiding paths 1 and 2 such that the mixing angles along those paths are
$\varphi_\eta$ and $\bar\varphi_{\eta}$ respectively. The braiding
transformations are given by $U$ and $\bar U$. We can now form a local mixing
operation $\bar U^\dagger U$ with mixing angles $\varphi_\eta-\bar\varphi_\eta$,
but since there is no local mixing by assumption we have
$\varphi_\eta-\bar\varphi_\eta=0$. Therefore the $\varphi_\eta$ are path
independent. The situation is illustrated in figure \eqref{fig:signargument}.

We now argue that $\varphi_1=-\varphi_2$. In our simple toy model this can be
checked by means of a straightforward calculation of $U_\text{local}$, but we will
give a more general argument instead.  In the way we decomposed the braiding
transformation, $U_{\text{local}}$ does not depend one the presence of the
second Kramers pair of MBS. We can therefore imagine moving the second Kramers
pair a little out of the way, such that the first Kramers pair can make a loop,
which does not encircle the second one, but passes through the initial position
of the second one. During the first half of this loop $\bm {X}_1$ will acquire a
mixing angle $\varphi_1$, which is  the same as the mixing angle it acquires
during an exchange. During the second half of this loop $\bm {X}_1$ will acquire
a mixing angle $\varphi_2$, which is the same as the mixing angle $\bm {X}_2$
acquires during an exchange.  In the absence of local mixing the overall mixing
angle for the loop is $\varphi_1+\varphi_2=0$. This situation is illustrated in
Fig. \ref{fig:signargument}.

We showed that in the absence of local mixing $\varphi_1=-\varphi_2$ in which
case we can bring Eq. \eqref{eqn:generalU} to the form of Eq.
\eqref{eqn:generalAfterBraidingBasisTransformation} by means of the local basis
transformation $W=e^{\frac {\varphi_1} 4 \chi_1\tilde{\chi}_1 -  \frac
{\varphi_1} 4 \chi_2\tilde{\chi}_2}$, such that
\begin{align}
    WUW^\dagger   &=e^{\frac \pi 4(\chi_1\chi_2 + \tilde{\chi}_1\tilde{\chi}_2)}.
    \label{eqn:independentBraiding}
\end{align}
This is only meaningful if the $\varphi_\eta$ and therefore $W$ do not depend on
the specific braiding path, which we also showed. The question arises what the
physical meaning of this particular basis choice is. This question can only be
answered meaningfully if we also specify a condition that ensures the absence of
local mixing. In our earlier work \cite{wolms_local_2014}, we specified
sufficient symmetry conditions that guaranteed the absence of local mixing and
in this case the basis choice leading to Eq.  \eqref{eqn:independentBraiding} is
the one respecting this additional symmetry.

It should be noted that a symmetry condition (or some other kind of condition)
that guarantees the absence of local mixing does not need to be global. It only
needs to be true locally around the Kramers pair. That means that one could try
to engineer a system such that the symmetry condition is approximately satisfied
close to the topological phase boundaries, where the Kramers pairs of MBS are
localized. If this remains true as the phase boundary and hence the Kramers pair
moves, than one can get path independent braiding statistics.

Even with the ideal transformation of Eq. \eqref{eqn:independentBraiding} the
question remains whether the braiding can yield something more interesting then
moving localized fermions around. The answer is yes, but it would require
schemes that initialize and read out non-local states. One can for instance
think of bringing two Kramers pairs of MBSs close together such that they split
in energy and one can initialize them in the lowest energy state of the split
states, after which one brings them apart again. Operations on those states can
then be performed similarly to the operations performed on class D Majorana
qubit systems. The non-local states in class DIII systems are actually entangled
with respect to the local bases. However, local mixing will decohere the
superpositions and therefore absence of local mixing is crucial for any
operations involving entanglement in the local basis.

In conclusion, we derived the most general braiding transformation for Kramers
pairs of Majorana  bound states. This transformation naturally decomposes into a
transformation that describes the independent exchange of two pairs of MBS and
additional  local rotations of the Kramers pairs that are being exchanged. The
angles of those local rotations are generally path dependent and independent for
the two Kramers pairs. The fact that the angles are path dependent shows again
that Kramers pairs of MBS cannot be used for topological quantum computation.
The derived form of the transformation is the most general one allowed by global
and local parity conservation and therefore the result is not limited to the
simple toy model, which we used to study it.  We also defined the concept of
local mixing, and showed that whether or not the phases are path dependent is
equivalent to the presence or absence of local mixing.  We also argued that in
the absence of local mixing the exchange transformation reduces to two
independent exchange of pairs of MBS without any path dependence. This is
helpful because it might be easier to check/engineer the presence or absence of
local mixing, because it is a local property.

We thank P. Brouwer for useful discussions. The Center for Quantum Devices
is funded by the Danish National Research Foundation. The research was supported
by The Danish Council for Independent Research \textbar{} Natural Sciences. A.
S. acknowledges support from the European Research Council (ERC), Minerva
foundation, the U.S.-Israel BSF, and Microsoft’s Station Q.

%


\begin{thebibliography}{15}%
\makeatletter
\providecommand \@ifxundefined [1]{%
 \@ifx{#1\undefined}
}%
\providecommand \@ifnum [1]{%
 \ifnum #1\expandafter \@firstoftwo
 \else \expandafter \@secondoftwo
 \fi
}%
\providecommand \@ifx [1]{%
 \ifx #1\expandafter \@firstoftwo
 \else \expandafter \@secondoftwo
 \fi
}%
\providecommand \natexlab [1]{#1}%
\providecommand \enquote  [1]{``#1''}%
\providecommand \bibnamefont  [1]{#1}%
\providecommand \bibfnamefont [1]{#1}%
\providecommand \citenamefont [1]{#1}%
\providecommand \href@noop [0]{\@secondoftwo}%
\providecommand \href [0]{\begingroup \@sanitize@url \@href}%
\providecommand \@href[1]{\@@startlink{#1}\@@href}%
\providecommand \@@href[1]{\endgroup#1\@@endlink}%
\providecommand \@sanitize@url [0]{\catcode `\\12\catcode `\$12\catcode
  `\&12\catcode `\#12\catcode `\^12\catcode `\_12\catcode `\%12\relax}%
\providecommand \@@startlink[1]{}%
\providecommand \@@endlink[0]{}%
\providecommand \url  [0]{\begingroup\@sanitize@url \@url }%
\providecommand \@url [1]{\endgroup\@href {#1}{\urlprefix }}%
\providecommand \urlprefix  [0]{URL }%
\providecommand \Eprint [0]{\href }%
\providecommand \doibase [0]{http://dx.doi.org/}%
\providecommand \selectlanguage [0]{\@gobble}%
\providecommand \bibinfo  [0]{\@secondoftwo}%
\providecommand \bibfield  [0]{\@secondoftwo}%
\providecommand \translation [1]{[#1]}%
\providecommand \BibitemOpen [0]{}%
\providecommand \bibitemStop [0]{}%
\providecommand \bibitemNoStop [0]{.\EOS\space}%
\providecommand \EOS [0]{\spacefactor3000\relax}%
\providecommand \BibitemShut  [1]{\csname bibitem#1\endcsname}%
\let\auto@bib@innerbib\@empty
\bibitem [{\citenamefont {Nayak}\ \emph {et~al.}(2008)\citenamefont {Nayak},
  \citenamefont {Simon}, \citenamefont {Stern}, \citenamefont {Freedman},\ and\
  \citenamefont {Das~Sarma}}]{nayak_non-abelian_2008}%
  \BibitemOpen
  \bibfield  {author} {\bibinfo {author} {\bibfnamefont {C.}~\bibnamefont
  {Nayak}}, \bibinfo {author} {\bibfnamefont {S.~H.}\ \bibnamefont {Simon}},
  \bibinfo {author} {\bibfnamefont {A.}~\bibnamefont {Stern}}, \bibinfo
  {author} {\bibfnamefont {M.}~\bibnamefont {Freedman}}, \ and\ \bibinfo
  {author} {\bibfnamefont {S.}~\bibnamefont {Das~Sarma}},\ }\href {\doibase
  10.1103/RevModPhys.80.1083} {\bibfield  {journal} {\bibinfo  {journal}
  {Reviews of Modern Physics}\ }\textbf {\bibinfo {volume} {80}},\ \bibinfo
  {pages} {1083} (\bibinfo {year} {2008})}\BibitemShut {NoStop}%
\bibitem [{\citenamefont {Beenakker}(2013)}]{beenakker_search_2013}%
  \BibitemOpen
  \bibfield  {author} {\bibinfo {author} {\bibfnamefont {C.}~\bibnamefont
  {Beenakker}},\ }\href {\doibase 10.1146/annurev-conmatphys-030212-184337}
  {\bibfield  {journal} {\bibinfo  {journal} {Annual Review of Condensed Matter
  Physics}\ }\textbf {\bibinfo {volume} {4}},\ \bibinfo {pages} {113} (\bibinfo
  {year} {2013})}\BibitemShut {NoStop}%
\bibitem [{\citenamefont {Leijnse}\ and\ \citenamefont
  {Flensberg}(2012)}]{leijnse_introduction_2012}%
  \BibitemOpen
  \bibfield  {author} {\bibinfo {author} {\bibfnamefont {M.}~\bibnamefont
  {Leijnse}}\ and\ \bibinfo {author} {\bibfnamefont {K.}~\bibnamefont
  {Flensberg}},\ }\href {\doibase 10.1088/0268-1242/27/12/124003} {\bibfield
  {journal} {\bibinfo  {journal} {Semiconductor Science and Technology}\
  }\textbf {\bibinfo {volume} {27}},\ \bibinfo {pages} {124003} (\bibinfo
  {year} {2012})}\BibitemShut {NoStop}%
\bibitem [{\citenamefont {Alicea}(2010)}]{alicea_majorana_2010}%
  \BibitemOpen
  \bibfield  {author} {\bibinfo {author} {\bibfnamefont {J.}~\bibnamefont
  {Alicea}},\ }\href {\doibase 10.1103/PhysRevB.81.125318} {\bibfield
  {journal} {\bibinfo  {journal} {Physical Review B}\ }\textbf {\bibinfo
  {volume} {81}},\ \bibinfo {pages} {125318} (\bibinfo {year}
  {2010})}\BibitemShut {NoStop}%
\bibitem [{\citenamefont {Wong}\ and\ \citenamefont
  {Law}(2012)}]{wong_majorana_2012}%
  \BibitemOpen
  \bibfield  {author} {\bibinfo {author} {\bibfnamefont {C.~L.~M.}\
  \bibnamefont {Wong}}\ and\ \bibinfo {author} {\bibfnamefont {K.~T.}\
  \bibnamefont {Law}},\ }\href {\doibase 10.1103/PhysRevB.86.184516} {\bibfield
   {journal} {\bibinfo  {journal} {Physical Review B}\ }\textbf {\bibinfo
  {volume} {86}},\ \bibinfo {pages} {184516} (\bibinfo {year}
  {2012})}\BibitemShut {NoStop}%
\bibitem [{\citenamefont {Zhang}\ \emph {et~al.}(2013)\citenamefont {Zhang},
  \citenamefont {Kane},\ and\ \citenamefont
  {Mele}}]{zhang_time-reversal-invariant_2013}%
  \BibitemOpen
  \bibfield  {author} {\bibinfo {author} {\bibfnamefont {F.}~\bibnamefont
  {Zhang}}, \bibinfo {author} {\bibfnamefont {C.~L.}\ \bibnamefont {Kane}}, \
  and\ \bibinfo {author} {\bibfnamefont {E.~J.}\ \bibnamefont {Mele}},\ }\href
  {\doibase 10.1103/PhysRevLett.111.056402} {\bibfield  {journal} {\bibinfo
  {journal} {Physical Review Letters}\ }\textbf {\bibinfo {volume} {111}},\
  \bibinfo {pages} {056402} (\bibinfo {year} {2013})}\BibitemShut {NoStop}%
\bibitem [{\citenamefont {Nakosai}\ \emph {et~al.}(2013)\citenamefont
  {Nakosai}, \citenamefont {Budich}, \citenamefont {Tanaka}, \citenamefont
  {Trauzettel},\ and\ \citenamefont {Nagaosa}}]{nakosai_majorana_2013}%
  \BibitemOpen
  \bibfield  {author} {\bibinfo {author} {\bibfnamefont {S.}~\bibnamefont
  {Nakosai}}, \bibinfo {author} {\bibfnamefont {J.~C.}\ \bibnamefont {Budich}},
  \bibinfo {author} {\bibfnamefont {Y.}~\bibnamefont {Tanaka}}, \bibinfo
  {author} {\bibfnamefont {B.}~\bibnamefont {Trauzettel}}, \ and\ \bibinfo
  {author} {\bibfnamefont {N.}~\bibnamefont {Nagaosa}},\ }\href {\doibase
  10.1103/PhysRevLett.110.117002} {\bibfield  {journal} {\bibinfo  {journal}
  {Physical Review Letters}\ }\textbf {\bibinfo {volume} {110}},\ \bibinfo
  {pages} {117002} (\bibinfo {year} {2013})}\BibitemShut {NoStop}%
\bibitem [{\citenamefont {Nakosai}\ \emph {et~al.}(2012)\citenamefont
  {Nakosai}, \citenamefont {Tanaka},\ and\ \citenamefont
  {Nagaosa}}]{nakosai_topological_2012}%
  \BibitemOpen
  \bibfield  {author} {\bibinfo {author} {\bibfnamefont {S.}~\bibnamefont
  {Nakosai}}, \bibinfo {author} {\bibfnamefont {Y.}~\bibnamefont {Tanaka}}, \
  and\ \bibinfo {author} {\bibfnamefont {N.}~\bibnamefont {Nagaosa}},\ }\href
  {\doibase 10.1103/PhysRevLett.108.147003} {\bibfield  {journal} {\bibinfo
  {journal} {Physical Review Letters}\ }\textbf {\bibinfo {volume} {108}},\
  \bibinfo {pages} {147003} (\bibinfo {year} {2012})}\BibitemShut {NoStop}%
\bibitem [{\citenamefont {Deng}\ \emph {et~al.}(2012)\citenamefont {Deng},
  \citenamefont {Viola},\ and\ \citenamefont {Ortiz}}]{deng_majorana_2012}%
  \BibitemOpen
  \bibfield  {author} {\bibinfo {author} {\bibfnamefont {S.}~\bibnamefont
  {Deng}}, \bibinfo {author} {\bibfnamefont {L.}~\bibnamefont {Viola}}, \ and\
  \bibinfo {author} {\bibfnamefont {G.}~\bibnamefont {Ortiz}},\ }\href
  {\doibase 10.1103/PhysRevLett.108.036803} {\bibfield  {journal} {\bibinfo
  {journal} {Physical Review Letters}\ }\textbf {\bibinfo {volume} {108}},\
  \bibinfo {pages} {036803} (\bibinfo {year} {2012})}\BibitemShut {NoStop}%
\bibitem [{\citenamefont {Keselman}\ \emph {et~al.}(2013)\citenamefont
  {Keselman}, \citenamefont {Fu}, \citenamefont {Stern},\ and\ \citenamefont
  {Berg}}]{keselman_inducing_2013}%
  \BibitemOpen
  \bibfield  {author} {\bibinfo {author} {\bibfnamefont {A.}~\bibnamefont
  {Keselman}}, \bibinfo {author} {\bibfnamefont {L.}~\bibnamefont {Fu}},
  \bibinfo {author} {\bibfnamefont {A.}~\bibnamefont {Stern}}, \ and\ \bibinfo
  {author} {\bibfnamefont {E.}~\bibnamefont {Berg}},\ }\href {\doibase
  10.1103/PhysRevLett.111.116402} {\bibfield  {journal} {\bibinfo  {journal}
  {Physical Review Letters}\ }\textbf {\bibinfo {volume} {111}},\ \bibinfo
  {pages} {116402} (\bibinfo {year} {2013})}\BibitemShut {NoStop}%
\bibitem [{\citenamefont {Gaidamauskas}\ \emph {et~al.}(2014)\citenamefont
  {Gaidamauskas}, \citenamefont {Paaske},\ and\ \citenamefont
  {Flensberg}}]{gaidamauskas_majorana_2014}%
  \BibitemOpen
  \bibfield  {author} {\bibinfo {author} {\bibfnamefont {E.}~\bibnamefont
  {Gaidamauskas}}, \bibinfo {author} {\bibfnamefont {J.}~\bibnamefont
  {Paaske}}, \ and\ \bibinfo {author} {\bibfnamefont {K.}~\bibnamefont
  {Flensberg}},\ }\href {\doibase 10.1103/PhysRevLett.112.126402} {\bibfield
  {journal} {\bibinfo  {journal} {Physical Review Letters}\ }\textbf {\bibinfo
  {volume} {112}},\ \bibinfo {pages} {126402} (\bibinfo {year}
  {2014})}\BibitemShut {NoStop}%
\bibitem [{\citenamefont {Haim}\ \emph {et~al.}(2014)\citenamefont {Haim},
  \citenamefont {Keselman}, \citenamefont {Berg},\ and\ \citenamefont
  {Oreg}}]{haim_time-reversal-invariant_2014}%
  \BibitemOpen
  \bibfield  {author} {\bibinfo {author} {\bibfnamefont {A.}~\bibnamefont
  {Haim}}, \bibinfo {author} {\bibfnamefont {A.}~\bibnamefont {Keselman}},
  \bibinfo {author} {\bibfnamefont {E.}~\bibnamefont {Berg}}, \ and\ \bibinfo
  {author} {\bibfnamefont {Y.}~\bibnamefont {Oreg}},\ }\href {\doibase
  10.1103/PhysRevB.89.220504} {\bibfield  {journal} {\bibinfo  {journal}
  {Physical Review B}\ }\textbf {\bibinfo {volume} {89}},\ \bibinfo {pages}
  {220504} (\bibinfo {year} {2014})}\BibitemShut {NoStop}%
\bibitem [{\citenamefont {Danon}\ and\ \citenamefont
  {Flensberg}(2015)}]{danon_interaction_2015}%
  \BibitemOpen
  \bibfield  {author} {\bibinfo {author} {\bibfnamefont {J.}~\bibnamefont
  {Danon}}\ and\ \bibinfo {author} {\bibfnamefont {K.}~\bibnamefont
  {Flensberg}},\ }\href {\doibase 10.1103/PhysRevB.91.165425} {\bibfield
  {journal} {\bibinfo  {journal} {Physical Review B}\ }\textbf {\bibinfo
  {volume} {91}},\ \bibinfo {pages} {165245} (\bibinfo {year}
  {2015})}\BibitemShut {NoStop}%
\bibitem [{\citenamefont {W{\"o}lms}\ \emph {et~al.}(2014)\citenamefont
  {W{\"o}lms}, \citenamefont {Stern},\ and\ \citenamefont
  {Flensberg}}]{wolms_local_2014}%
  \BibitemOpen
  \bibfield  {author} {\bibinfo {author} {\bibfnamefont {K.}~\bibnamefont
  {W{\"o}lms}}, \bibinfo {author} {\bibfnamefont {A.}~\bibnamefont {Stern}}, \
  and\ \bibinfo {author} {\bibfnamefont {K.}~\bibnamefont {Flensberg}},\ }\href
  {\doibase 10.1103/PhysRevLett.113.246401} {\bibfield  {journal} {\bibinfo
  {journal} {Physical Review Letters}\ }\textbf {\bibinfo {volume} {113}},\
  \bibinfo {pages} {246401} (\bibinfo {year} {2014})}\BibitemShut {NoStop}%
\bibitem [{\citenamefont {Sau}\ \emph {et~al.}(2011)\citenamefont {Sau},
  \citenamefont {Clarke},\ and\ \citenamefont {Tewari}}]{sau_controlling_2011}%
  \BibitemOpen
  \bibfield  {author} {\bibinfo {author} {\bibfnamefont {J.~D.}\ \bibnamefont
  {Sau}}, \bibinfo {author} {\bibfnamefont {D.~J.}\ \bibnamefont {Clarke}}, \
  and\ \bibinfo {author} {\bibfnamefont {S.}~\bibnamefont {Tewari}},\ }\href
  {\doibase 10.1103/PhysRevB.84.094505} {\bibfield  {journal} {\bibinfo
  {journal} {Physical Review B}\ }\textbf {\bibinfo {volume} {84}},\ \bibinfo
  {pages} {094505} (\bibinfo {year} {2011})}\BibitemShut {NoStop}%
\end{thebibliography}

\end{document}